\documentclass[letterpaper, 10 pt, conference]{ieeeconf}  

\IEEEoverridecommandlockouts                              

\let\proof\relax

\usepackage[printwatermark]{xwatermark}
\usepackage{xcolor}
\usepackage{multirow}
\usepackage{amsmath}
\usepackage{amsthm}
\usepackage{amssymb}
\usepackage{graphicx} 
\usepackage{subfig}
\usepackage{booktabs}
\newtheorem{theorem}{Theorem}

\newtheorem{proposition}[theorem]{Proposition}

\newcommand{\ie}{\textit{i}.\textit{e}. }

\theoremstyle{definition}
\newcommand{\ra}[1]{\renewcommand{\arraystretch}{#1}}

\title{\LARGE \bf
Extended Dynamic Mode Decomposition with Learned Koopman Eigenfunctions for Prediction and Control
}

\author{Carl Folkestad$^{*,1}$, Daniel Pastor$^{*,1}$, Igor Mezic$^{^3}$, Ryan Mohr$^{^2}$, Maria Fonoberova$^{^2}$, and Joel Burdick$^{1}$\thanks{*Both authors contributed equally, $^{1}$ Division of Engineering and Applied Sciences, California Institute of Technology, Pasadena, CA 91125, USA, $^{2}$ AIMDyn Inc., $^{3}$ University of California, Santa Barbara} \thanks{Corresponding authors: \texttt{carl.folkestad@caltech.edu, dpastorm@caltech.edu}}}

\begin{document}
\maketitle
\thispagestyle{empty}
\pagestyle{empty}

\begin{abstract}
This paper presents a novel learning framework to construct Koopman eigenfunctions for unknown, nonlinear dynamics using data gathered from experiments. The learning framework can extract spectral information from the full nonlinear dynamics by learning the eigenvalues and eigenfunctions of the associated Koopman operator. We then exploit the learned Koopman eigenfunctions to learn a lifted linear state-space model. To the best of our knowledge, our method is the first to utilize Koopman eigenfunctions as lifting functions for EDMD-based methods. We demonstrate the performance of the framework in state prediction and closed loop trajectory tracking of a simulated cart pole system. Our method is able to significantly improve the controller performance while relying on linear control methods to do nonlinear control. 
\end{abstract}

\section{Introduction}
A key step in developing a high performance robotic application is the modeling of the robot's mechanics. Standard modelling and identification require extensive knowledge of the system and laborious system identification procedures~\cite{lupashin2014platform}. Moreover, although methods to show stability and safety of nonlinear systems exist \cite{Khalil2002, Ames2017}, control design incorporating state and control limitations remains challenging.

{\em Learning} can capture the salient aspects of a robot's complex mechanics and environmental interactions.  Gaussian process dynamical systems models \cite{GaussianDynamic} can identify nonlinear affine control models in a non-parametric way. Alternatively, spectrally normalized neural networks \cite{Shi2018} can fit dynamics models with stability guarantees. Yet, effective nonlinear control design incorporating state and actuator constraints after identifying the model can be challenging. Deep neural networks for control Lyapunov function augmentation \cite{Taylor2019a} can be used for control design with different types of constraints but learns a task-specific augmentation that cannot be used for other objectives. Similarly, model-free reinforcement learning (MFRL) \cite{Duan2016} learns feedback policies that implicitly incorporate the robot's dynamics. However, sample efficiency is very low.  Moreover, while safety during MFRL is now possible \cite{Garcia2015, Cheng:AAAI2019}, one cannot yet guarantee that learned policies will satisfy performance requirements or state and actuator limits.

Our work contributes to Koopman inspired modelling and identification techniques, which have received substantial recent attention \cite{Rowley2009, Budisic2012}. In particular, the Dynamic Mode Decomposition (DMD) and extended DMD (EDMD) methods have emerged as efficient numerical algorithms to identify finite dimensional approximations of the Koopman operator associated with the system dynamics \cite{Schmid2010, Williams2015}. The methods are easy to implement, mainly relying on least squares regression, and computationally and mathematically flexible, enabling numerous extensions and applications \cite{Brunton2019}.

For example, DMD-based methods have been successfully used in the field of fluid mechanics to capture low-dimensional structure in complex flows \cite{Taira2017}, in robotics for external perturbation force detection \cite{Berger2015}, and in neuroscience to identify dynamically relevant features in ECOG data \cite{Brunton2016}. More recently, Koopman-style modeling has been extended to {\em controlled} nonlinear systems \cite{Kaiser, Proctor2018}. This is particularly interesting as EDMD can be used to approximate nonlinear control systems by a lifted state space model. As a result, well developed linear control design methods such as robust, adaptive, and model predictive control (MPC) \cite{Korda2018a} can be utilized to design nonlinear controllers.


Typically, EDMD-methods employ a dictionary of functions used to lift the state variables to a space where the dynamics are approximately linear. However, if not chosen carefully, the time evolution of the dictionary functions cannot be described by a linear combination of the other functions in the dictionary. This results in error accumulation when the model is used for prediction, potentially causing significant prediction performance degradation. To mitigate this we develop a learning framework that can extract spectral information from the full nonlinear dynamics by learning the eigenvalues and eigenfunctions of the associated Koopman operator. Limited attention has been given to constructing eigenfunctions from data. Sparse identification techniques have been used to identify approximate eigenfunctions \cite{Kaiser2017} but rely on defining an appropriate candidate function library. Other previous methods (e.g., \cite{Korda2018}) depend upon assumptions that are problematical for robotic systems: the ID data is gathered while the robot operates under open loop controls, which can lead to catastrophic system damage.

This paper presents a novel learning framework, \textit{Koopman Eigenfunction Extended Dynamic Mode Decomposition} (KEEDMD), to construct Koopman eigenfunctions for unknown, nonlinear dynamics using a data gathered from experiments. We then exploit the learned Koopman eigenfunctions to learn a lifted linear state-space model. To the best of our knowledge, our method is the first to utilize Koopman eigenfunctions as lifting functions for EDMD-based methods. Furthermore, we demonstrate that the identified model can readily be used with MPC \cite{Mayne2000} on simulated experiments.

\subsection{Notation} 
We denote the space of all continuous functions on some domain $\mathcal{X} \subset \mathbb{R}^d$ as $\mathcal{C}(\mathcal{X})$, the Jacobian of the function $f(\mathbf{x})$ evaluated at $\mathbf{x}=\mathbf{a}$ is denoted $\mathbf{D}f(a)$. $\mathbb{N}_0$ is the set of natural numbers including zero. $I$ is the identity matrix of appropriate dimensions. $\delta_{jk}$ is the kronecker delta, $\delta_{jk} = 1$ if and only if $j=k$.

\section{Preliminaries on Koopman Operator Theory}\label{sec:preliminaries}
This section briefly reviews basic facts about the Koopman operator, and then summarizes key results that form the theoretical underpinnings for the Koopman eigenfunction learning methodology presented in Section \ref{sec:eigenfunctions}.

\subsection{The Koopman Operator}
Consider the autonomous dynamical system:
  \begin{equation}\label{eq:aut_dynamics}
    \mathbf{\dot{x}} = f(\mathbf{x}) = A\mathbf{x} + v(\mathbf{x})
  \end{equation}
with state $\mathbf{x} \in \mathcal{X} \subset \mathbb{R}^d$ and $f(\cdot)$ Lipschitz continuous on $\mathcal{X}$. We assume that system (\ref{eq:aut_dynamics}) has a fixed point at the origin: $f(0) = 0$. For a system with a single attractor in $\mathcal{X}$ this can be achieved without loss of generality by a change of coordinates. The flow of this dynamical system is denoted by $S_t(\mathbf{x})$ and is defined as
\begin{equation}
    \frac{d}{dt}S_t(\mathbf{x}) = f(S_t(\mathbf{x}))
\end{equation}
for all $\mathbf{x} \in \mathcal{X}$ and all $t\geq 0$. The \textit{Koopman operator semi-group} $(U_t)_{t\geq0}$, hereafter denoted as the \textit{Koopman operator}, is defined as 
\begin{equation}
    U_t \gamma = \gamma \circ S_t
\end{equation}
for all $\gamma \in \mathcal{C}(\mathcal{X})$, where $\circ$ denotes function composition. Each element of the Koopman operator maps continuous functions to continuous functions, $U_t: \mathcal{C}(\mathcal{X}) \rightarrow \mathcal{C}(\mathcal{X})$. Crucially, each $U_t$ is a \textit{linear} operator. An \textit{eigenfunction} of the Koopman operator associated to an eigenvalue $e^{\lambda} \in \mathbb{C}$ is any function $\phi \in \mathcal{C}(\mathcal{X})$ that defines a coordinate evolving linearly along the flow of (\ref{eq:aut_dynamics}) satisfying
\begin{equation}\label{eq:koopman_eig}
    (U_t \phi)(\mathbf{x}) = \phi(S_t(\mathbf{x})) = e^{\lambda t} \phi (\mathbf{x})
\end{equation}

\subsection{Construction of Eigenfunctions for Nonlinear Dynamics}\label{sec:prelim_eigfunc}

For any sufficiently smooth autonomous dynamical system that is asymptotically stable to a fixed point, Koopman eigenfunctions can be constructed by first finding the eigenfunctions of the system linearization around the fixed point and then composing them with a diffeomorphism \cite{Mohr2014}. To see this, consider asymptotically stable dynamics of the form (\ref{eq:aut_dynamics}).  The linearization of the dynamics around the origin is
\begin{equation}\label{eq:lin_dynamics}
 \mathbf{\dot{y}} = \mathbf{D}f(0)\mathbf{y} = \hat{A}\mathbf{y}, \text{ } \mathbf{y} \in \mathcal{Y}
\end{equation}
The following proposition describes how to construct eigenfunction-eigenvalue pairs
for the linearized system (\ref{eq:lin_dynamics}).
\begin{proposition}\label{prop:eigfunc}
Let  $\hat{A}_1$ denote the linearization (\ref{eq:lin_dynamics}) of the nonlinear system (\ref{eq:aut_dynamics}) with $\mathcal{Y}$ scaled into the unit hypercube, $\mathcal{Y}_1 \subset \mathcal{Q}_1$, and let $\{\mathbf{v}_1, \dots, \mathbf{v}_d\}$ be a basis of the eigenvectors of $\hat{A}_1$  corresponding to nonzero eigenvalues $\{\lambda_1, \dots, \lambda_d\}$. Let $\{\mathbf{w}_1, \dots, \mathbf{w}_d\}$ be the adjoint basis to $\{\mathbf{v}_1, \dots, \mathbf{v}_d\}$ such that $\langle \mathbf{v}_j,\mathbf{w}_k \rangle = \delta_{jk}$ and $\mathbf{w}_j$ is an eigenvector of $\hat{A}_1^*$ at eigenvalue $\Bar{\lambda}_j$. Then, the linear functional
\begin{equation} \label{eq:lin_eigfunc}
    \psi_j(\mathbf{y}) = \langle \mathbf{y}, \mathbf{w}_j \rangle
\end{equation}
\noindent is a nonzero eigenfunction of $U_{\hat{A}_1}$, the Koopman operator associated to $\hat{A}_1$. 
Furthermore, for any tuple $(m_1, \dots, m_d) \in \mathbb{N}_0^d$
\begin{equation}\label{eq:eig_products}
    \bigg ( \prod_{j=1}^d e^{m_j \lambda_j}, \prod_{j=1}^d \psi_j^{m_j} \bigg )
\end{equation}{}
is an eigenpair of the Koopman operator $U_{\hat{A}_1}.$
\end{proposition}
\proof{
A less formal description of the results in the proposition and associated proofs are described in \cite{Mohr2014}, Example 4.6. By utilizing inner-product properties, $\psi_j$ is an eigenfunction of $U_{\hat{A}}$ as described in (\ref{eq:koopman_eig}) since 
\begin{align*}
    (U_t\psi_j)(\mathbf{y}) &= U_t \langle \mathbf{y}, \mathbf{w}_j \rangle = \langle \mathbf{y}, U_t^*\mathbf{w}_j \rangle = \langle \mathbf{y}, \Bar{e^{\lambda_j}}\mathbf{w}_j \rangle \\
    &= e^{\lambda_j} \langle \mathbf{y}, \mathbf{w}_j \rangle = e^{\lambda_j}\psi_j(\mathbf{y})
\end{align*} 
 \noindent By scaling the state-space such that $\mathcal{Y}_1 \subset \mathcal{Q}_1$, the linear eigenfunctions (\ref{eq:lin_eigfunc}) form a vector space on $\mathcal{Y}_1$ that is closed under point-wise products. The construction of arbitrarily many eigenpairs (\ref{eq:eig_products}) therefore follows from the semi-group property of eigenfunctions (see \cite{Budisic2012}, Prop. 5). \qed 
}

In the following we denote the linear functionals (\ref{eq:lin_eigfunc}) as \textit{principal eigenfunctions}. The eigenfunctions for the Koopman operator associated with the linearized dynamics can be used to construct eigenfunctions associated with the Koopman operator of the nonlinear dynamics through the use of a {\em conjugacy map}, as described in the following proposition.

\begin{proposition} \label{prop:conjugacy}
Assume that the nonlinear system (\ref{eq:aut_dynamics}) is topologically conjugate to the linearized system (\ref{eq:lin_dynamics}) via the diffeomorphism $h:\mathcal{X} \rightarrow \mathcal{Y}$. Let $B \in \mathcal{X}$ be a simply connected, bounded, positively invariant open set in $\mathcal{X}$ such that $h(B) \subset Q_r \subset \mathcal{Y}$, where $Q_r$ is a cube in $\mathcal{Y}$. Scaling $Q_r$ to the unit cube $Q_1$ via the smooth diffeomorphism $g: Q_r \rightarrow Q_1$ gives $(g \circ h)(B) \subset Q_1$. Then, if $\psi$ is an eigenfunction for $U_{\hat{A}_1}$ at $e^{\lambda}$, then $\psi \circ g \circ h$ is an eigenfunction for $U_f$ at eigenvalue $e^{\lambda}$, where $U_f$ is the Koopman operator associated with the nonlinear dynamics (\ref{eq:aut_dynamics}).
\end{proposition}{}
\proof{
See \cite{Budisic2012}, Proposition 7. \qed
}

\noindent The following extension of the Hartman-Grobman theorem guarantees the existence of the diffeomorphism, $h$ described in Proposition \ref{prop:conjugacy}, between the linearized and nonlinear systems in the entire basin of attraction of a fixed point, for sufficiently smooth dynamics.

\begin{theorem}\label{thm:diffeo}
Consider the system (\ref{eq:aut_dynamics}) with $v(\mathbf{x}) \in \mathcal{C}^2(\mathcal{X})$. Assume that matrix $A\in \mathbb{R}^{d\times d}$ is Hurwitz, i.e., all of its eigenvalues have negative real parts. So, the fixed point $\mathbf{x} = \mathbf{0}$ is exponentially stable and let $\Omega$ be its basin of attraction. Then $\exists h(\mathbf{x}) \in \mathcal{C}^1(\Omega):\Omega \rightarrow \mathbb{R}^d$, such that 
  \begin{equation}\label{eq:diffeo}
      \mathbf{y} = c(\mathbf{x})=\mathbf{x}+h(\mathbf{x})
  \end{equation} 
is a $\mathcal{C}^1$ diffeomorphism with $\mathbf{D}c(\mathbf{0})=I$ in $\Omega$ and satisfies $\mathbf{\dot{y}} = A \mathbf{y}$.
\end{theorem}{}
\proof{
See \cite{Lan2013}, Theorem 2.3. \qed
}

\subsection{Koopman Theory for Controlled Systems}
There are several ways to extend the Koopman operator to actuated systems such that systems with external forcing can be analyzed through the spectral properties of its associated Koopman operator \cite{Korda2018a, Proctor2018}. These observations underpins the adaption of EDMD methods to controlled systems to construct finite-dimensional approximations to the Koopman operator. In particular, given a dictionary of D dictionary functions $\phi(x)$ and $N$ data snapshots of the states, X, control inputs, U, and state derivatives, Y, from a $n$-dimensional system with $m$ control inputs, a linear regression problem can be formulated as

\begin{equation}
    \min_{A \in \mathbb{R}^{(D \times D)},B \in {(D \times m)}} ||A \phi(X) + B U - Y||
\end{equation}{}

This results in a linear model of the dynamics of the form $\mathbf{\dot{z}} = A \mathbf{z} + B \mathbf{u}$ where the outputs of interest are predicted by $y = C\mathbf{z}$ where $C$ can be approximated by another regression problem aiming to minimize $||CZ - Y||$ \cite{Korda2018a}.

\section{Motivating Analytic Example}
Certain systems have a structure that leads to a closed Koopman subspace if a correct set of observables is chosen. In this section, we demonstrate how the theory presented in Section \ref{sec:preliminaries} can be used to construct eigenfunctions when the system dynamics are known and we can analytically construct the diffeomorphism described in Theorem 3. We consider the system

\begin{equation} \label{eq:example_dynamics}
    \begin{bmatrix}\dot{x}_1\\\dot{x}_2\end{bmatrix} = \begin{bmatrix} \mu x_1 \\ \lambda(x_2 - x_1^2)\end{bmatrix} .
\end{equation}{}

which has a finite dimensional Koopman operator. This is used in Section \ref{sec:example_koop} to construct three eigenfunctions that can completely describe the evolution of the system by utilizing the Koopman modes associated with each eigenfunction \cite{Budisic2012}. Then, we demonstrate how to arrive at the same eigenfunctions through the use of the diffeomorphism in Section \ref{sec:example_diff}. This underpins our data-driven approach described in Section \ref{sec:eigenfunctions}, using data to approximate the conjugacy map when the dynamics are unknown and/or a exact diffeomorphism cannot be derived.

\subsection{Calculating Eigenfunctions from the Koopman Operator} \label{sec:example_koop}

By choosing observables $y = [x_1, x_2, x_1^2]^T$, (\ref{eq:example_dynamics}) can be rewritten as an equivalent linear system
\begin{equation}
    \begin{bmatrix}\dot{y}_1\\ \dot{y}_2 \\ \dot{y}_3\end{bmatrix} = \underbrace{\begin{bmatrix} \mu & 0 & 0 \\ 0 & \lambda & -\lambda \\ 0 & 0 &2 \mu \end{bmatrix}}_K \begin{bmatrix}y_1\\ y_2 \\ y_3\end{bmatrix}
\end{equation}{}
where K is the Koopman operator of the system. From this we can construct three Koopman eigenfunctions of (\ref{eq:example_dynamics}). Let $\{\mathbf{v}_i\}_{i=1}^3$ be the eigenvectors of $K$ and let $\{\mathbf{w}_i\}_{i=1}^3$ be the adjoint basis to $\{\mathbf{v}_i\}_{i=1}^3$ scaled such that $\langle \mathbf{w}_i, \mathbf{v}_j \rangle = \delta_{ij}$. Then, three eigenfunctions of the system are
\begin{align} \label{eq:example_analytic_eig}
\begin{split}
    &\psi_1(\mathbf{y}) = \langle \mathbf{y}, \mathbf{w}_1 \rangle = y_1 = x_1\\
    &\psi_2(\mathbf{y}) = \langle \mathbf{y}, \mathbf{w}_2 \rangle = y_3 = x_1^2\\
    &\psi_3(\mathbf{y}) = \langle \mathbf{y}, \mathbf{w}_3 \rangle = y_2 + \frac{\lambda}{\lambda - 2 \mu}y_3 = x_2 + \frac{\lambda}{\lambda - 2 \mu}x_1^2
\end{split}{}
\end{align}{}

\subsection{Calculating Eigenfunctions Based on the Diffeomorphism} \label{sec:example_diff}

We now show how the calculated eigenfunctions can be obtained through the diffeomorphism between the linearized and nonlinear dynamics. The linearization of the dynamics (\ref{eq:example_dynamics}) around the origin is
\begin{equation}
    \begin{bmatrix}\dot{\hat{x}}_1\\ \dot{\hat{x}}_2\end{bmatrix} = \underbrace{\begin{bmatrix}\mu & 0 \\ 0 & \lambda \end{bmatrix}}_A \begin{bmatrix}\hat{x}_1\\ \hat{x}_2 \end{bmatrix}
\end{equation}{}
and we can construct principal eigenfunctions for the linearized system, $\hat{\psi}_1(\mathbf{x}) = \langle \mathbf{\hat{w}}_1, \mathbf{x} \rangle = x_1$, $\hat{\psi}_2(\mathbf{x}) = \langle \mathbf{\hat{w}}_2, \mathbf{x} \rangle = x_2$, where $\mathbf{\hat{w}}_1, \mathbf{\hat{w}}_2$ are the eigenvectors of the adjoint of $A$. As described in Proposition 1, we can construct arbitrarily many eigenfunctions for the linearized system by taking powers and products of the principal eigenfunctions, i.e. $\hat{\psi}_i(\mathbf{x}) = \hat{\psi}^{m_i^{(1)}}_1(\mathbf{x})\hat{\psi}^{m_i^{(2)}}_2(\mathbf{x}) =  x_1^{m_i^{(1)}}x_2^{m_i^{(2)}}$ is an eigenfunction of the linearized system. 

To get the eigenfunctions for the nonlinear system, it can be shown that 
\begin{equation}
    c(\mathbf{x}) = \begin{bmatrix} x_1 \\ x_2\end{bmatrix} + \begin{bmatrix} 0 \\ \frac{\lambda}{\lambda - 2 \mu}x_1^2 \end{bmatrix}
\end{equation}{}
is a diffeomorphism of the form described in Theorem 3. Then, ignoring the scaling function $g(\mathbf{x})$ for simplicity of exposition, we get the following eigenfunctions for the nonlinear dynamics
\begin{align}
\begin{split}
    &\phi_1(\mathbf{x}) = \hat{\psi}_1(c(\mathbf{x})) = x_1 \\
    &\phi_2(\mathbf{x}) = \hat{\psi}_2(c(\mathbf{x})) = x_2 + \frac{\lambda}{\lambda - 2 \mu}x_1^2 \\
    &\phi_i(\mathbf{x}) = \hat{\psi}_i(c(\mathbf{x})) = x_1^{m^{(1)}_i} \bigg(x_2 + \frac{\lambda}{\lambda - 2 \mu}x_1^2 \bigg )^{m^{(2)}_i}, \, i = 3, \dots 
\end{split}{}
\end{align}{}
and we see that with $m_3 = (2, 0)$ we recover the analytic eigenfunctions of Equation (\ref{eq:example_analytic_eig}). 

\section{Data-driven Koopman Eigenfunctions for Unknown Nonlinear Dynamics} \label{sec:eigenfunctions}
We now develop the data-driven approach to learn the diffeomorphism $h(x)$ described in Proposition 2 and Equation~\ref{eq:diffeo}, resulting in a methodology for constructing Koopman eigenfunctions from data. 

    \subsection{Modeling Assumptions}
    We consider the dynamical system
    \begin{equation}\label{eq:true_dynamics}
    \mathbf{\dot{x}} = a(\mathbf{x}) + B \mathbf{u}
    \end{equation}
    \noindent where $\mathbf{x} \in \mathcal{X} \subset \mathbb{R}^d,\text{ } a(\mathbf{x}): \mathcal{X} \rightarrow \mathcal{X},\text{ }\mathbf{u} \in \mathcal{U} \subset \mathbb{R}^m,\text{ }B \in \mathbb{R}^{d \times m}$, and where $a(\mathbf{x})$ and $B$ are unknown. We assume that we have access to a nominal linear model
    \begin{equation} \label{eq:nom_dynamics}
        \mathbf{\dot{x}} = A_{nom}\mathbf{x} + B_{nom}\mathbf{u}
    \end{equation}
    \noindent where $\mathbf{x} \in \Omega \subset \mathcal{X} \subset \mathbb{R}^d, \text{ }A_{nom} \in \mathbb{R}^{(d\times d)}, \text{ }B_{nom} \in \mathbb{R}^{(d \times m)}, \text{ } \mathbf{u} \in \mathcal{U}$ and an associated nominal linear feedback controller $\mathbf{u}^{nom} = K_{nom}\mathbf{x}$ that stabilizes the system (\ref{eq:true_dynamics}) to the origin in a region of attraction $\Omega$ around the origin. The nominal model (\ref{eq:nom_dynamics}) can for example be obtained from first principles modeling or from parameter identification techniques and linearization of the constructed model around the fixed point if needed.
    
    \subsection{Constructing Eigenfunctions from Data}
    
    Algorithm 1 constructs Koopman eigenfunctions from data, based on the foundations introduced in Section \ref{sec:prelim_eigfunc}. $M_t$ trajectories of fixed length $T$ are executed from initial conditions $\mathbf{x}_0^j \in \Omega \text{ } j=1,\dots,M_t$, and are guided by the nominal control law $\mathbf{u}^{nom}$. The system's states and control actions are sampled at a fixed interval $\Delta t$, resulting in a data set 
      \begin{equation} 
         \mathcal{D} = \left(\left(\mathbf{x}_k^j, \mathbf{u}_k^j\right)_{k=0}^{M_s}\right)_{j=1}^{M_t}
      \end{equation}
    where $M_s = T/\Delta t$. Variable length trajectories and sampling rates can be implemented with minor modifications. 
    \begin{figure}[t] 
        \vspace{0.11cm}
        \centering
        \includegraphics[width=0.48\textwidth]{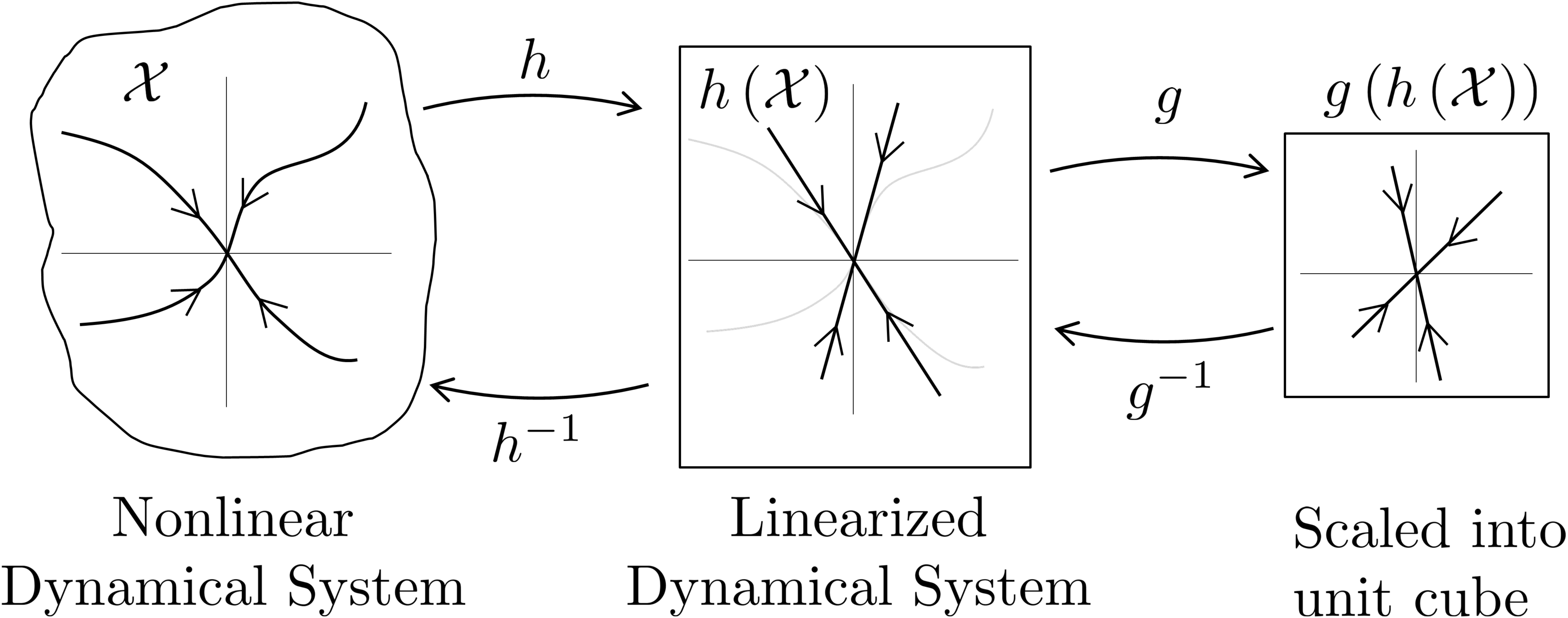}
        \caption{Chain of topological conjugacies used to construct eigenfunctions, adapted from \cite{Mohr2014}.}
        \label{fig:h_diagram}
    \end{figure}

    \begin{figure}[b]
    \centering
    \small
    \begin{tabular}{p{8cm}}
         \hline
         \textbf{Algorithm 1} Data-driven Koopman Eigenpair Construction\\
         \hline
         \textbf{Require:} Data set $\mathcal{D} = \big((\mathbf{x}_k^j, \mathbf{u}_k^j)_{k=0}^{M_s}\big)_{j=1}^{M_t}$, nominal model matrices $A_{nom}$, $B_{nom}$, nominal control gains $K_{nom}$, number of lifting functions $N$, $N$ power combinations $(m_1^{(i)},\dots,m_d^{(i)})\in\mathbb{N}_0^d, i = 1,\dots,N$
    \end{tabular}
    \begin{tabular}{p{0.1cm} p{7.5cm}}
    \\
        1: & Construct principal eigenpairs for the linearized dynamics: $(\lambda_j, \psi_j(\mathbf{y})) \leftarrow{} (\lambda_j, \langle \mathbf{y}, \mathbf{w}_j \rangle), \qquad j=1,\dots,n$\\
        2: & Construct $N$ eigenpairs from the principal eigenpairs: $(\Tilde{\lambda}_i, \Tilde{\psi}_i) \leftarrow \big ( \sum_{j=1}^d \lambda_j^{m_j^{(i)}}, \prod_{j=1}^d \psi_j^{m_j^{(i)}} \big ),\quad i=1,\dots,N$\\
        3: & Fit diffeomorphism estimator: $h(\mathbf{y}) \leftarrow \text{ERM}(\mathcal{H}_h, \mathcal{L}_h, \mathcal{D})$\\
        4: & Construct scaling function: $g(\mathbf{y}) \leftarrow g: \mathcal{Q}_r \rightarrow \mathcal{Q}_1$\\
        5: & Construct $N$ eigenpairs for the nonlinear dynamics: $(\Tilde{\lambda}_i, \phi_i) \leftarrow (\Tilde{\lambda}_i, \Tilde{\psi}_i(g(c(\mathbf{x})))), \qquad i=1,\dots,N$
    \end{tabular}
    \begin{tabular}{p{8cm}}
    \\
         \textbf{Output:} $\Lambda = \text{diag}(\Tilde{\lambda}_1,\dots,\Tilde{\lambda}_N), \qquad \boldsymbol{\phi} = [\phi_1,\dots,\phi_N]^T$\\
         \hline
    \end{tabular}
    \end{figure}
    
    Under the nominal control law, Koopman eigenfunctions for the nominal linearized model (\ref{eq:nom_dynamics}) can be constructed as in Proposition \ref{prop:eigfunc} using the eigenvectors and eigenvalues of the closed loop dynamics matrix $A_{cl} = A_{nom}+B_{nom}K_{nom}$. I.e. let $\mathcal{Q}_r$ be a hypercube of radius $r$ such that $\mathcal{X} \subset \mathcal{Q}_r$, a scaling function $g: \mathcal{Q}_r \rightarrow \mathcal{Q}_1$ can then be constructed (by scaling each coordinate) to get the scaled dynamics matrix $A_{cl,1}$. Furthermore, let $\{\mathbf{v_j}\}_{j=1}^d$ be a basis of eigenvectors of $A_{cl,1}$ with corresponding eigenvalues $\{\lambda_j\}_{j=1}^d$ and let $\{\mathbf{w_j}\}_{j=1}^d$ be the adjoint basis to $\{\mathbf{v_j}\}_{j=1}^d$. Then $\psi_j(\mathbf{y}) = \langle \mathbf{y}, \mathbf{w}_j \rangle$ is an eigenfunction of $U_{A_{cl,1}}$ with eigenvalue $e^{\lambda}_j$ and we can construct an arbitrary number of eigenpairs using the product rule (\ref{eq:eig_products}).
    
    The eigenfunction construction for the linearized system only relies on the nominal model. To construct Koopman eigenfunctions for the true nonlinear dynamical system, we aim to learn the diffeomorphism (\ref{eq:diffeo}) between the linearized model (\ref{eq:nom_dynamics}) and the true dynamics (\ref{eq:true_dynamics}), see Figure \ref{fig:h_diagram}. This diffeomorphism is guaranteed to exist in the entire basin of attraction $\Omega$ by Theorem 3. Let $\mathcal{H}_h$ be a class of continuous nonlinear function mapping $\mathbb{R}^d$ to $\mathbb{R}^d$. The diffeomorphism is found by solving the following optimization problem:
    \begin{align}\label{eq:diff_opt}
    \begin{split}
        \min_{h \in \mathcal{H}_h} \sum_{k=1}^{M_t} \sum_{j=1}^{M_s} & (\mathbf{\dot{x}}_k^j + \dot{h}(\mathbf{x}_k^j) - A_{cl}(\mathbf{x}_k^j + h(\mathbf{x}_k^j)))^2\\
        \text{s.t.} \qquad & \mathbf{D}h(\mathbf{0}) = \mathbf{0}
    \end{split}
    \end{align}
    \begin{figure*}[!t]
    \normalsize
    \begin{subequations} \label{eq:EDMD_losses}
    \begin{alignat}{4}
    &\text{\textbf{Position dynamics:}} \qquad \qquad \,\, \min_{B_\mathbf{p} \in \mathbb{R}^{(d/2)\times m}} &&||\mathbf{y}_\mathbf{p} - X_\mathbf{p}B_\mathbf{p}^T||_2^2, &&X_\mathbf{p} = [U], \quad &&\mathbf{y}_\mathbf{p} = [\dot{P} - I V]
    \\
    &\text{\textbf{Velocity dynamics:}} \quad \min_{A_\mathbf{v} \in \mathbb{R}^{(d/2) \times (n+N)}, B_\mathbf{v} \in \mathbb{R}^{(d/2)\times m}}  && ||\mathbf{y}_\mathbf{v} - X_\mathbf{v}[A_\mathbf{v} \,\, B_\mathbf{v}]^T||_2^2, 
    \quad && X_\mathbf{v} = [P \,\, V \,\, \Phi \,\, U], \,\,\,\,\,&&\mathbf{y}_\mathbf{v} = [\dot{V}]
    \\
    &\text{\textbf{Eigenfunction dynamics:}}
                     \qquad \,\,\, \min_{B_{\boldsymbol{\phi}} \in \mathbb{R}^{N\times m}} \quad &&||\mathbf{y}_{\boldsymbol{\phi}} - X_{\boldsymbol{\phi}}B_{\boldsymbol{\phi}}^T||_2^2, 
                    &&X_{\boldsymbol{\phi}} = [U - U_{nom}], && \mathbf{y}_{\boldsymbol{\phi}} = [\dot{\Phi}-\Lambda \Phi]
    \end{alignat}{}
    \end{subequations}{}
    \vspace{-0.9cm}
    \end{figure*}

    %
    \noindent which is a direct transformation of Theorem \ref{thm:diffeo} into the setting with unknown nonlinear dynamics. The form of problem (\ref{eq:diff_opt}) is found by minimizing the squared loss  $\mathbf{\dot{y}}_k-A_{cl}\mathbf{y}_k$ over all data pairs, substituting  $\mathbf{y} = \mathbf{x} + h(\mathbf{x})$, and adding the constraint $\mathbf{D}c(\mathbf{0}) = I$ results in the formulated optimization problem (\ref{eq:diff_opt}).
    
    We next formulate (\ref{eq:diff_opt}) as a general supervised learning problem. Consider the data set of input-output pairs $\mathcal{D}_h = \big \{ (\mathbf{x}_k, \mathbf{\dot{x}}_k), \mathbf{\dot{x}}_k-A_{cl}\mathbf{x}_k \big \}_{k=1}^{M_s \cdot M_t}$, constructed from the state measurements (perhaps by calculating  numerical derivatives $\mathbf{\dot{x}}_k^j$ as needed), and aggregated to a data matrix. The class $\mathcal{H}_h$ can be any function class suitable for supervised learning (e.g. deep neural networks) as long as the Jacobian of the function $h(\mathbf{x}) \in \mathcal{H}_h$ w.r.t. the input can be readily calculated. Assuming $h(\mathbf{x}) \in \mathcal{H}_h$ we define the loss function
    \begin{align}
    \begin{split}
        &\mathcal{L}_h(\mathbf{x},\mathbf{\dot{x}}, A_{cl}\mathbf{x}-\mathbf{\dot{x}}) = \\
        & \qquad \, || \dot{h}(\mathbf{x}) - A_{cl}h(\mathbf{x}) - (A_{cl}\mathbf{x} - \mathbf{\dot{x}}) ||^2 + \alpha||\mathbf{D}h(\mathbf{0})||^2\\
        & \quad  = || \mathbf{D} h(\mathbf{x})\mathbf{\dot{x}} - A_{cl}h(\mathbf{x}) - (A_{cl}\mathbf{x} - \mathbf{\dot{x}}) ||^2 + \alpha||\mathbf{D}h(\mathbf{0})||^2
    \end{split}
    \end{align}
    where parameter $\alpha$ penalizes the violation of constraint  (\ref{eq:diff_opt}). The supervised learning goal is to select a function in $\mathcal{H}_h$ through empirical risk minimization (ERM):
    \begin{equation} \label{eq:erm}
        \min_{h \in \mathcal{H}_h} \frac{1}{M_s \cdot M_t} \sum_{k=1}^{M_s \cdot M_t} \mathcal{L}_h(\mathbf{x}_k,\mathbf{\dot{x}}_k, A_{cl}\mathbf{x}_k-\mathbf{\dot{x}}_k)\ .
    \end{equation}

    Finally, with function $h$ identified from ERM (\ref{eq:erm}), Proposition \ref{prop:conjugacy} implies that the Koopman eigenfunctions for the unknown dynamics under the nominal control law can be constructed from the eigenfunctions of the linearized system by the function composition:
    \begin{equation}\label{eq:nonlin_eigfunc}
    \phi_j(\mathbf{x}) = \Tilde{\psi}_j(g(h(\mathbf{x})))
    \end{equation}
    \noindent where $g$ is the scaling function ensuring that the basin of attraction $\Omega$ is scaled to lie within the unit hypercube $Q_1$ and $\Tilde{\psi}_j$ is an eigenfunction for the linearized system with associated eigenvalue $\Tilde{\lambda}_j$ constructed with (\ref{eq:eig_products}). 
    
    Importantly, because the diffeomorphism is learned from data, it may not perfectly capture the underlying diffeomorphism over all of $\Omega$, and thus the eigenfunctions for the unknown dynamics are approximate. The error arises from the fact that the ERM problem is underdetermined resulting in the possibility of multiple approximations with equal loss while failing to capture the underlying diffeomorphism. This is especially an issue when encountering states and state time derivatives not reflected in the training data and introduces a demand for exploratory control inputs to cover a larger region of the state space of interest. This can be achieved by introducing a random perturbation of the control action deployed on the system and is akin to persistence of excitation in adaptive control \cite{Ljung1987}. To understand these effects, state dependent model error bounds are needed but they are out of the scope of this paper.
    
    \section{Koopman Eigenfunction Extended Dynamic Mode Decomposition}\label{sec:KEEDMD}
    
    To use the constructed Koopman eigenfunctions for prediction and control, we develop an EDMD-based method to build a linear model in a lifted space. Since this method exploits the structure of the Koopman eigenfunctions, it is dubbed {\em Koopman Eigenfunction Extended Dynamic Mode Decomposition} (KEEDMD). We construct $N$ eigenfunctions $\{\phi_j\}_{j=1}^N$ with associated eigenvalues $\Lambda = \text{diag}(\lambda_1,\dots,\lambda_N)$ as outlined in Section \ref{sec:eigenfunctions} and define the lifted state as 
       \begin{equation} \label{eq:lifted_state} 
           \mathbf{z} = [\mathbf{x}, \boldsymbol{\phi}(\mathbf{x})]^T
       \end{equation}
    where $\boldsymbol{\phi}(\mathbf{\mathbf{x}}) = [\phi_1(\mathbf{x}), \dots, \phi_N(\mathbf{x})]$. We seek to learn a model of the form 
       \begin{equation}\label{eq:lifted_equation}
          \mathbf{\dot{z}} = A\mathbf{z} + B \mathbf{u}
       \end{equation}
    where matrices $A \in \mathbb{R}^{(N+d)\times(N+d)}, B\in \mathbb{R}^{(N+d) \times m}$ are unknown, and are to be inferred from the collected data. 
    
    We focus on systems governed by Lagrangian dynamics, whose state space coordinates consist of position, $\mathbf{p}$, and velocity $\mathbf{v}$:  $\mathbf{x} = [\mathbf{p}, \mathbf{v}]^T$, with $\mathbf{\dot{p}} = \mathbf{v}$. The rows of $A$ corresponding to the position states are known. Furthermore, by construction the eigenvalues $\Lambda$ describe the evolution of the eigenfunctions under the nominal control law. Therefore, the rows of $A$ corresponding to eigenfunctions are also known. As a result, the lifted state space model has the following structure:
    \begin{equation}\label{eq:KEEDMD}
    \begin{bmatrix} \mathbf{\dot{p}} \\ \mathbf{\dot{v}} \\ \boldsymbol{\dot{\phi}} \bigg(\!\!\begin{bmatrix} \mathbf{p} \\ \mathbf{v} \end{bmatrix}\!\!\bigg) \end{bmatrix} =\underbrace{\begin{bmatrix} \begin{array}{cc} 0 & \quad I \end{array}{} & 0\\
    \begin{array}{cc} A_{\mathbf{v}\mathbf{p}}& A_{\mathbf{v}\mathbf{v}} \end{array}{}&A_{\mathbf{v}\boldsymbol{\phi}} \\
    -B_{\boldsymbol{\phi}}K_{nom} & \begin{array}{c}  \\  \end{array}{} \Lambda \begin{array}{c} \\ \end{array}{} \end{bmatrix}}_A
    \begin{bmatrix} \mathbf{p} \\ \mathbf{v} \\ \boldsymbol{\phi} \bigg(\!\!\begin{bmatrix} \mathbf{p} \\ \mathbf{v} \end{bmatrix}\!\!\bigg) \end{bmatrix}
    +\underbrace{\begin{bmatrix} B_\mathbf{p} \\
    B_\mathbf{v}\\
    B_{\boldsymbol{\phi}} \begin{array}{c}  \\  \end{array}{} \end{bmatrix}}_B  \mathbf{u}
    \end{equation}
    \noindent where $0, I, \Lambda, K_{nom}$ are fixed matrices and $A_{\mathbf{v}\mathbf{p}}, A_{\mathbf{v}\mathbf{v}}, A_{\mathbf{v}\boldsymbol{\phi}},\\ B_{\mathbf{p}}, B_{\mathbf{v}}, B_{\boldsymbol{\phi}}$ are determined from data. The term $-B_{\boldsymbol{\phi}} K_{nom}$ accounts for the effect of the nominal controller on the evolution of the eigenfunctions. To infer the different parts of (\ref{eq:KEEDMD}), we construct the data matrices and formulate the loss function for three separate ordinary least squares regression problems defined in Equation (\ref{eq:EDMD_losses}). The data matrices are aggregations of the data snapshots, e.g. $P = [\mathbf{p}_1^1,\dots,\mathbf{p}_{M_s}^1,\dots,\mathbf{p}_1^{M_t},\dots,\mathbf{p}_{M_s}^{M_t}]^T$. Furthermore, $P, V, \Phi, U, U_{nom}$ are derived from measurements and $\dot{P}, \dot{V}, \dot{\Phi}$ are found by numerically differentiating $P, V, \Phi$, respectively. $U$ and $U_{nom}$ are related by $U = U_{nom} + U_{pert}$, where $U_{nom}$ is the nominal linear feedback control action and $U_{pert}$ is the added random perturbation to induce exploratory behavior as discussed in Section \ref{sec:eigenfunctions}. The KEEDMD exploits the control perturbation to learn the effect of actuation on the Koopman eigenfunctions.
    
    To reduce overfitting, different forms of regularization can be added to the objectives of the regression formulations. In particular, LASSO-regularization promoting sparsity in the learned matrices has been shown to perform well for dynamical systems \cite{Brunton} when used in normal EDMD. This has also been the case in our numerical simulation, where LASSO-regularization seem to improve the prediction performance and the stability of the results.
    
    When the lifted state space model is identified, state estimates can be obtained as $\mathbf{x} = C \mathbf{z}$, where $C = [I \quad 0]$. $C$ is denoted the \textit{projection matrix} of the lifted state space model. 





\subsection{Extensions for Trajectory-tracking Nominal Controller}
In all of the above, a pure state feedback nominal control law is considered. We now discuss how to extend the methodology to allow linear trajectory-tracking feedback controllers of the form $\mathbf{u} = K_{nom}(\mathbf{x}-\boldsymbol{\tau}(t))$. Under this controller, the closed loop linearized dynamics become $\dot{\mathbf{x}} = A_{nom}\mathbf{x} + B_{nom}K_{nom}(\mathbf{x} - \boldsymbol{\tau}(t))$. Let the definition of the closed loop dynamics matrix, $A_{cl} = (A_{nom}+B_{nom}K_{nom})$ and the principal eigenfunctions (the eigenfunctions associated with the Koopman operator of the linearized system), be as in Section \ref{sec:prelim_eigfunc}. Then, the evolution of the principal eigenfunctions becomes

\begin{align}
\begin{split}
    \dot{\psi}_j(\mathbf{y}) = \dot{\langle \mathbf{w}_j, \mathbf{y} \rangle} &= \mathbf{w}_j^T \dot{\mathbf{y}} \\
    &= \langle \mathbf{w}_j, A_{cl}\mathbf{y} - B_{nom}K_{nom}\boldsymbol{\tau}(t)\rangle\\
    &= \lambda_j \langle \mathbf{w}_j, \mathbf{y} \rangle - \langle \mathbf{w}_j, B_{nom}K_{nom}\boldsymbol{\tau}(t) \rangle \\
    &= \lambda_j \psi(\mathbf{y}) - \mathbf{w}_j^T B_{nom}K_{nom}\boldsymbol{\tau}(t)
\end{split}{}
\end{align}{}

where $\lambda_j$ and $\mathbf{w}_j$ are the j-th eigenvalue and adjoint eigenvector of $A_{cl}$, respectively. Notably, the principal eigenfunctions  evolves as described in Section \ref{sec:eigenfunctions} but with an additional forcing term, $-\mathbf{w}_j^T B_{nom}K_{nom}\boldsymbol{\tau}(t)$.

Utilizing that the dynamics considered have linear actuated dynamics (see Eq. \ref{eq:true_dynamics}), we show that the evolution of the eigenfunctions of the Koopman operator associated with the full dynamics is affine in the input signal.

\begin{proposition}
Assume that $B_{nom}$ in the linearized model of the dynamics (\ref{eq:nom_dynamics}) is equal to the actuation matrix of the true dynamics (\ref{eq:true_dynamics}) and that the dynamics are controlled by a linear trajectory-tracking feedback controller of the form $u=K_{nom}(\mathbf{x}-\boldsymbol{\tau}(t)$. Then, the time derivatives of the eigenfunctions of the Koopman operator associated with the dynamics (\ref{eq:true_dynamics}) constructed as described in Proposition 1-2 are affine in the external forcing signal $\boldsymbol{\tau}(t)$.
\end{proposition}{}

\begin{proof}
We first show that the diffeomorphism between the linearized and nonlinear dynamics is linear in the forcing signal. Consider the diffeomorphism described in Theorem 3 with an additional forcing term. Derived from the linearized dynamics, we seek to find $h(\mathbf{x})$ such that
\begin{equation}\label{eq:traj_diff}
    \dot{\mathbf{y}} = A_{cl}\mathbf{y} - B_{nom}K_{nom}\boldsymbol{\tau}(t), \qquad \mathbf{y} = \mathbf{x} + h(\mathbf{x})
\end{equation}{}

By algebraic manipulations we get that
\begin{align}
    \begin{split}
        \dot{\mathbf{y}} &= \dot{\mathbf{x}} + \dot{h}(\mathbf{x}) = A_{cl}(\mathbf{x} + h(\mathbf{x}) - B_{nom}K_{nom}\boldsymbol{\tau}(t)\\
        &\Rightarrow a(x) + BK_{nom}(\mathbf{x}-\boldsymbol{\tau}(t)) + \dot{h}(\mathbf{x}) \\
        &\,\,= (A_{nom}+B_{nom}K_{nom})(\mathbf{x} + h(\mathbf{x})) - B_{nom}K_{nom}\boldsymbol{\tau}(t)\\
        &\Rightarrow\dot{h}(\mathbf{x}) - A_{cl}h(\mathbf{x}) = A_{nom}\mathbf{x} - a(\mathbf{x})
    \end{split}{}
\end{align}{}
Hence, $h(\mathbf{x})$ does not depend on the forcing signal $\boldsymbol{\tau}(t)$. As a result, the diffeomorphism $c(\mathbf{x})$ does not depend on the forcing signal and the eigenfunctions associated with the eigenfunctions of the nonlinear dynamics (\ref{eq:true_dynamics}) evolve affinely in the forcing signal.
\end{proof}{}

Because the eigenfunctions evolve linearly in the forcing signal, the KEEDMD-framework can readily learn the effect of external forcing on the eigenfunctions by minor modifications. First, the loss of the diffeomorphism empirical risk minimization (\ref{eq:diff_opt}) must be changed to account for the forcing term following the construction of (\ref{eq:traj_diff}) such that the new loss function becomes

\begin{align}
\begin{split}
    &\mathcal{L}_h(\mathbf{x},\mathbf{\dot{x}}, A_{cl}\mathbf{x}-\mathbf{\dot{x}}, \boldsymbol{\tau}(t)) = \\
    & \qquad \, || \dot{h}(\mathbf{x}) - A_{cl}h(\mathbf{x}) - (A_{cl}\mathbf{x} - \mathbf{\dot{x}}) + B_{nom}K_{nom}\boldsymbol{\tau}||^2 \\
    & \qquad + \alpha||\mathbf{D}h(\mathbf{0})||^2\\
\end{split}
\end{align}
where $\boldsymbol{\tau}$ is the vector of desired states corresponding to the time that $\mathbf{x}, \dot{\mathbf{x}}$ were sampled. Second, the data matrix $X_{\boldsymbol{\phi}}$ in the regression formulation (\ref{eq:EDMD_losses}) must be modified so the effect of the forcing on the eigenfunction evolution can be learned. This is achieved by setting $X_{\boldsymbol{\phi}} = [U - K_{nom}[P \quad V]]$.

\begin{figure*}[t] 
    \centering
    \subfloat[Open loop prediction]{\includegraphics[width=0.48\textwidth]{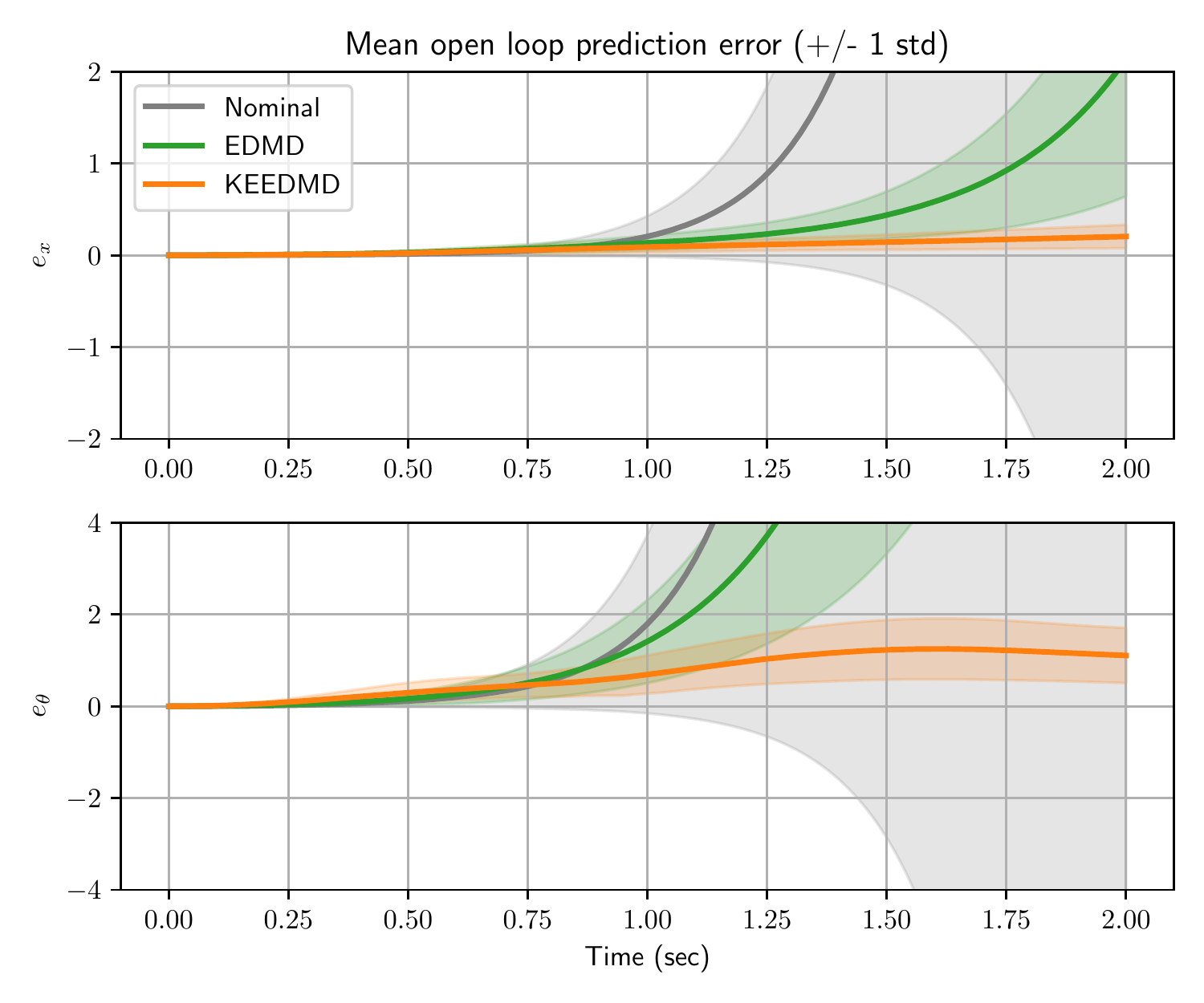}\label{fig:openloop}}
    \subfloat[Closed loop trajectory tracking]{\includegraphics[width=0.48\textwidth]{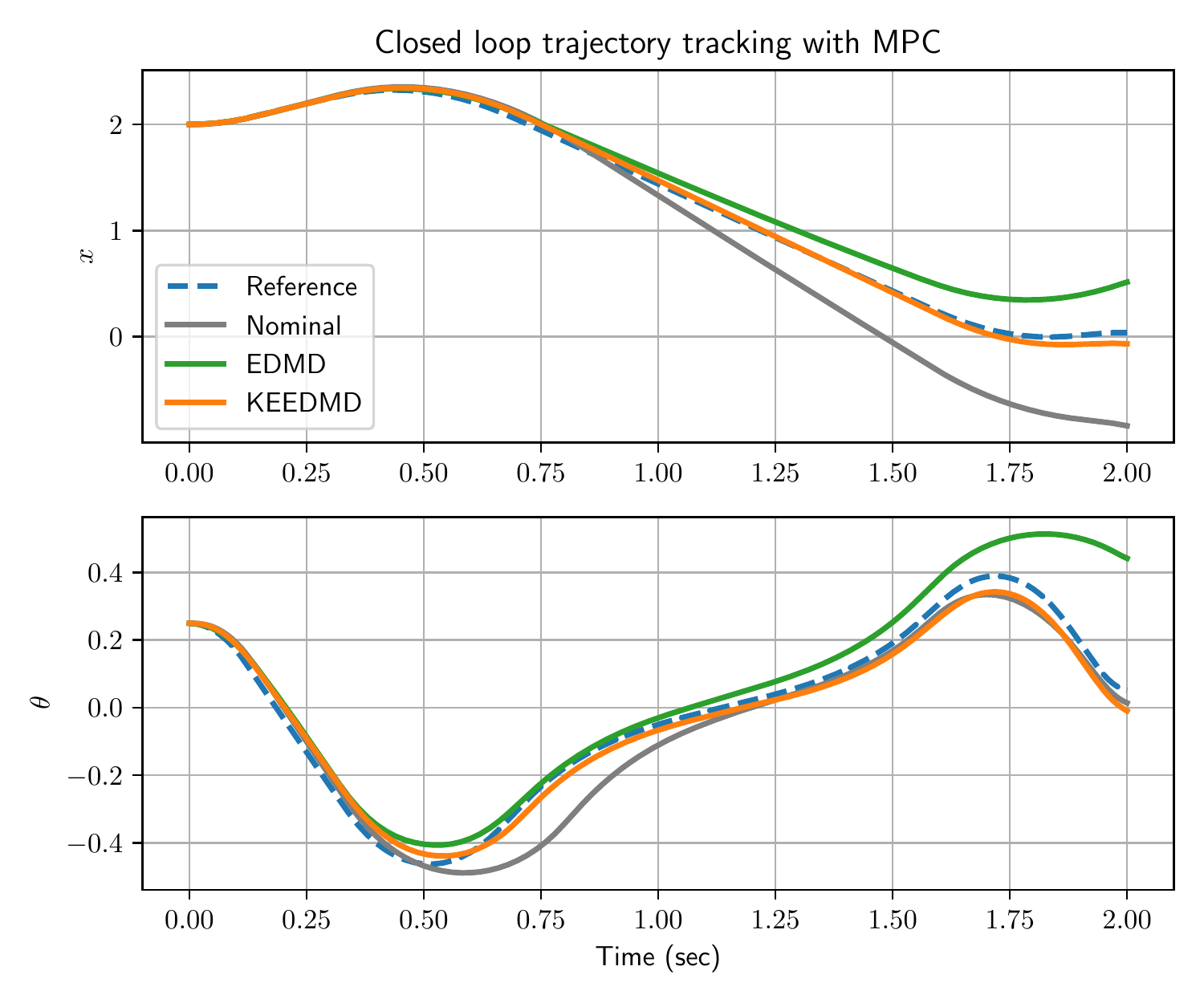}\label{fig:closedloop}} 
    \caption{Performance comparison of the nominal model, EDMD, and KEEDMD for (a) prediction and (b) closed loop.}
    \label{fig:prediction}
    \vspace{-0.5cm}
\end{figure*}

\section{Model Predictive Control Design}
Inspired by \cite{Korda2018a}, we use the Koopman operator to transform the original nonlinear optimization problem into an efficient quadratic program (QP) that is solved at each time step. The QP formulation requires us to discretize the previously learned linear continuous dynamics. We assume a known objective function that is solely a function of states and controls. For simplicity, we use a quadratic objective function but other objective functions are possible by adding them to the lifting functions. We assume known control bounds $u_{\min}, u_{\max} \in \mathbb{R}^{m}$ and state bounds $x_{\min}, x_{\max} \in \mathbb{R}^{n}$. These assumptions define the following optimization problem:

\begin{equation}\label{eq:ZMPC}
\begin{array}{ll}{\underset{\substack{u \in \mathbb{R}^{m \times N_p} \\ z\in \mathbb{R}^{N\times N_p}}}{\operatorname{min}}} & {\sum_{p=1}^{N_p}\left[\left(Cz_{p}-\tau_p\right)^{T} Q\left(Cz_{p}-\tau_p\right)+u_{p}^{T} R u_{p}\right]} \\ {\text{s.t.}} & {z_{p}=A_dz_{p-1}+B_du_{p}} \\ {} & {x_{\min } \leq Cz_{p} \leq x_{\max }} \,\,\,\,\,\,\,\,\,\,\,\,\,\,\,\, p=1,\dots,N_p\\ {} & {u_{\min } \leq u_{p} \leq u_{\max }} \\ {} & {z_{0}=\phi\left(x_k\right)}\end{array}
\end{equation}
here $Q \in \mathbb{R}^{n \times n}$ and $ R \in \mathbb{R}^{m \times m}$ are positive semidefinite cost matrices, $\tau \in \mathbb{R}^{n \times N_p}$ is the reference trajectory, $A_d \in \mathbb{R}^{N\times N}$ and $B_d\in \mathbb{R}^{N\times m}$ are the discrete time versions of (\ref{eq:lifted_equation}), $C \in \mathbb{R}^{n\times N}$ is the projection matrix, and $\phi \in \mathbb{R}^{N}$ are the eigenfunctions.
\\
To remove the dependency on the lifting dimension $N$ in Eq. (\ref{eq:ZMPC}), the state is eliminated via an explicit relation with the control input. This formulation is referred as the \textit{dense} form MPC. This step greatly reduces the number of optimization variables, which is beneficial as we must solve the MPC problem in real-time. In this form, the MPC is agnostic not only of the lifting dimension but of the whole Koopman formalism, \ie the eigenfunctions $\phi$ and linear matrices $A_d$, $B_d$ and $C$ do not directly appear in the formulation. 


\section{Experimental Results}

To obtain an initial evaluation of the performance of the proposed framework, we study the canonical cart pole system with continuous dynamics\footnote{The code for learning and control is publicly available on \texttt{https://github.com/Cafolkes/keedmd}}:

\begin{equation}
    \begin{bmatrix}{}\Ddot{x} \\ \Ddot{\theta}\end{bmatrix} = 
    \begin{bmatrix}{} \frac{1}{M+m} \big (m l \Ddot{\theta}\cos{\theta} - m l \dot{\theta}^2 + F \big)\\ \frac{1}{l}\big ( g\sin{\theta} + \Ddot{x}\cos{\theta}\big )  \end{bmatrix}
\end{equation}

\noindent where $x, \theta$ are the cart's horizontal position and the angle between the pole and the vertical axis, respectively, $M, m$ are the cart's and pole tip's mass, respectively, $l$ is the pole length, $g$ the gravitational acceleration, and $F$ the horizontal force input on the cart. The linearization of the dynamics around the origin is used as the nominal model. Starting with knowledge of the nominal model only, our goal is to learn a lifted state space model of the dynamics to improve the system's ability to track a trajectory designed based on the nominal model to move to the origin from a initial condition two meters away. 
We will collect data with a nominal controller, learn the lifted state space model and use this model to design an improved MPC.

To build the dataset used for training, 40 trajectories are simulated by sampling an initial point in the interval $(x, \theta, \dot{x}, \dot{\theta)} \in [-2.5,2.5] \times [-0.25,0.25] \times [-0.05,0.05] \times [-0.05,0.05]$, generating a two second long trajectory from the initial point to the origin with a MPC based on the nominal model, and simulating the system with a PD controller stabilizing the system to the trajectory. Note that the system is underactuated and stabilizing the system to a set point under PD control will not work. The PD controller is perturbed with white noise of variance 0.5 to aid the model fitting as described in Section \ref{sec:eigenfunctions} and state and control action snapshots are sampled from the simulated trajectories at 100 hz. With the collected data, eigenfunctions are constructed as described in Algorithm 1 and a lifted state space model is identified according to (\ref{eq:EDMD_losses}).

To benchmark our results, we compare our prediction and control results against (1) the nominal model, and (2) a EDMD-model with the state and Gaussian radial basis functions as lifting functions. In both the EDMD and KEEDMD models, a lifting dimension of 85 is used and elastic net regularization is added with regularization parameters determined by cross validation. The diffeomorphism, $h$, is parameterized by a 3-layer neural network with 50 units in each layer and implemented with \textit{PyTorch} \cite{Paszke}. The EDMD and KEEDMD regressions are implemented with \textit{Scikit-learn} \cite{PedregosaFABIANPEDREGOSA2011}. 

First, we compare the open loop prediction performance by generating sampling 40 points from the same intervals as the training data and then stabilizing the system to the origin with a MPC based on the nominal model with a 2 second prediction horizon. Then, the time evolution of the system is predicted from the sampled initial point and with the control sequence from the collected data for each trajectory with the nominal model, EDMD-model, and the KEEDMD-model. The mean  error between the predicted evolution and the true system evolution over all the trajectories is depicted in Figure \ref{fig:openloop}. Both the nominal and EDMD model is able to predict the evolution for the first second but den diverges. In contrast, KEEDMD is able to maintain good prediction performance over the entire duration of the trajectories with relatively low, constant standard deviation.

\begin{table}[b]
\ra{1.2}
\vspace{-0.3cm}
  \centering
  \caption{Improvement in MPC cost with learned models}
    \begin{tabular}{@{}lcc@{}}
    \toprule
    & \parbox[b]{2.8cm}{\centering \textbf{Improvement over}\\\textbf{nominal model}} & \parbox[b]{2.8cm}{\centering \textbf{Improvement over}\\\textbf{EDMD-model}}\\ 
    \midrule
    EDMD& $- 68.00\%$ &\\
    KEEDMD& $- 96.75 \%$ & $- 89.84 \%$\\
    \bottomrule
    \end{tabular}%
  \label{tab:mpc_cost}%
  \vspace{-0.2cm}
\end{table}%

To evaluate the closed loop performance, we compare the behaviour of the three different models on the task of moving from initial point ($x_0, \theta_0, \dot{x}_0, \dot{\theta}_0) = (2, 0.25, 0, 0)$ in two seconds. The nominal model is used to generate a trajectory from the initial point to the origin. Then, a dense form MPC using the learned lifted state space model is implemented in Python using the QP solver \textit{OSQP} \cite{Stellato2018}. The MPC costs on the trajectory tracking task are significantly improved when the lifted state space models are used, see Figure \ref{fig:closedloop}. It is important to note that the EDMD based MPC regulates less towards the end of the trajectory causing large deviations but still outperforms the nominal model in terms of MPC cost by 62 percent. For the same penalty matrices $Q,R$, the KEEDMD based MPC has significantly better trajectory tracking performance and further reduces the MPC cost by 90 percent.

\section{Conclusions and Future Work}
We presented a novel method to learn non-linear dynamics, using Koopman Eigenfunctions constructed from principal eigenfunctions and a non-linear diffeomorphism as lifting functions for Extended Dynamic Mode Decomposition (EDMD). We then used a MPC framework to obtain an optimal controller, while respecting state and control input bounds. We showed in simulation that the method drastically outperforms the linearization around the origin as well as the classical EDMD method with the same number of lifting functions in both prediction and closed loop control. These preliminary results show focusing on the spectral properties of the Koopman Operator allow for a more compact representation while achieving similar performance. Furthermore, we demonstrate that our methodology can be used to implement a nonlinear MPC in a highly computationally efficient manner by exploiting the linear structure and eliminating the dependence on the lifting dimension. In future work, this method will be applied on experimental platforms. Two of the main current limitations are that the proposed method does not allow state-dependent $B$ matrix in the true dynamics and the need to collect data using a linear stabilizing controller and current work is investigating these issues \cite{icra20}. In addition, different control strategies like robust and adaptive control can be investigated. 

\section*{Acknowledgement}
The authors would like to thank the four anonymous referees for their thoughtful comments that helped improve this manuscript. This work has been supported in part by Raytheon Company and the DARPA Physics of Artificial Intelligence program, HR00111890033. The first author is grateful for the support of the Aker Scholarship Foundation.


\bibliography{references,references_outside_mendeley} 
\bibliographystyle{ieeetr}

\end{document}